# ROBUST MORPHOLOGICAL MEASURES
# FOR LARGE–SCALE STRUCTURE

Thomas Buchert

*Max–Planck–Institut für Astrophysik, 85740 Garching, F. R. G.*

ABSTRACT

A complete family of statistical descriptors for the morphology of large–scale structure based on Minkowski–Functionals is presented. These robust and significant measures can be used to characterize the local and global morphology of spatial patterns formed by a coverage of point sets which represent galaxy samples. Basic properties of these measures are highlighted and their relation to the 'genus statistics' is discussed. Test models like a Poissonian point process and samples generated from a Voronoi–model are put into perspective.

## 1. Introduction

There is considerable effort in the community directed towards new statistical tools for the measurement of galaxy samples or cosmological model realizations (see the review articles in: Feigelson & Babu 1992). Many of these innovations proved useful, each of them puts emphasis on another aspect of the statistical characterization of structures, but none of them can be viewed as part of a concept for a *complete characterization* of the morphology of spatial patterns. Certainly, such concepts exist, like the familiar hierarchy of n–point correlation functions, but the realization of the full hierarchy is practically impossible.

We may be none the wiser than we were a couple of decades ago, but we certainly are better equipped: There is a well–developed mathematical literature on stochastic geometry and image analysis providing technologies with which we are able to accomplish the full characterization of morphology, provided we allow us to run for a while in strictly mathematical territory (see, e.g., Serra 1982). Of those technologies, we recently proposed the application of *Minkowski–Functionals* for the description of content, shape and connectivity of large–scale structure (Mecke *et al.* 1994).

Minkowski–Functionals provide a complete family of morphological measures, which are robust for small samples, independent of statistical assumptions on how point sets (like galaxy samples) are generated, and yield global as well as local morphological information. Moreover, their significance outweighs most of what can be achieved by other methods, like, e.g., the related 'genus statistics' (see: Melott 1990 for a review, and Section 3).

As an example of a non–robust measure which is not sensitive to the morphology of structures we look at the *two–point correlation function* (compare Buchert & Martínez 1993). Fig.1 presents two point processes which have been generated by a Voronoi–model of large–scale structure (van de Weygaert, priv. comm.). Their two–point correlation function as fitted by a power–law in a given spatial range is identical by construction. On one hand we appreciate by eye differences in the morphology of these point sets, especially on larger scales; on the other hand, these sets are quite similar so that we clearly need a good method to distinguish them.

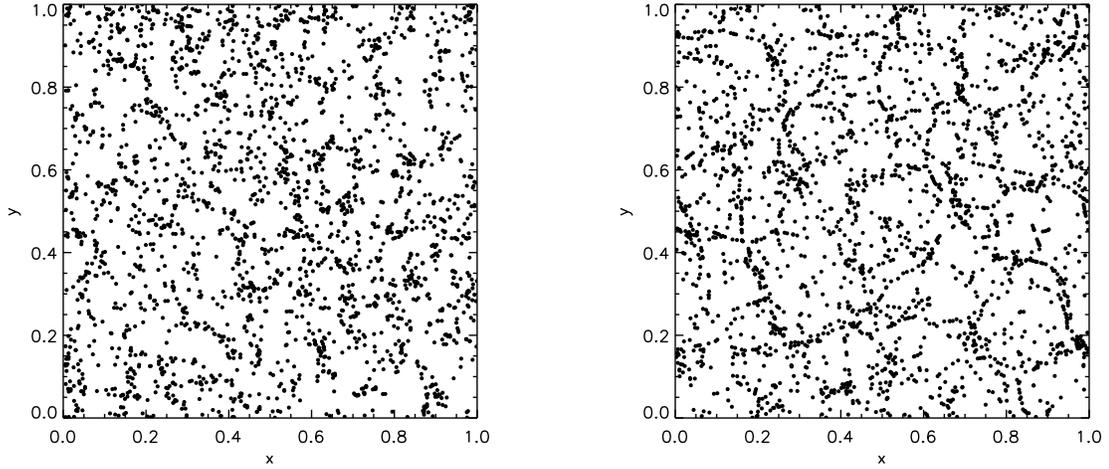

Fig. 1. Two point processes generated by a Voronoi–model are shown (projection of 1/4 of a cube with 10000 points). The two–point correlation functions of both point sets are identical by construction.

## 2. Minkowski–Functionals

### 2.1. A complete family of morphological descriptors

In order to calculate morphological characteristics of a point set, we mark each point in the point set $\{x_i, i = 1 \ldots N\}$ with a spherical ball $B_r(x_i)$ with a given radius $r$ (in general we could use an arbitrary convex grain). We form the *coverage* consisting of the union of balls $\mathcal{B}(r) = \cup_{i=1}^{N} B_r(x_i)$ and measure the resulting body with integral-geometrical methods: We calculate 1. the total volume, 2. the total surface area, 3. the integral mean curvature, and 4. the integral Gaussian curvature of the coverage. In three dimensions, these 4 familiar measures comprise the set of 4 Minkowski–Functionals $W_{\alpha, \alpha=0,1,2,3}$ of a body $\mathcal{K}$:

$$W_0 = V(\mathcal{K}) \; ; \; 3W_1 = F(\mathcal{K}) \; ; \; 3W_2 = H(\mathcal{K}) \; ; \; 3W_3 = G(\mathcal{K}) = 4\pi\chi(\mathcal{K}) \; . \qquad (1)$$

The integral Gaussian curvature $G$ is known as the *Euler–characteristic* $\chi$ in algebraic topology; $\chi = 1$ for a compact, convex body.

The noticable simplicity of these descriptors on one hand, has, on the other hand, a wide mathematical background in stochastic geometry. As an example I mention a theorem by Hadwiger which states, that *any* statistical measure with the following three properties can be represented as a linear combination of Minkowski–Functionals: 1. *Additivity* states that the measures can be calculated additively from their local values, e.g., functionals $W_\alpha$ of the union of two bodies, $\mathcal{K}_1$ and $\mathcal{K}_2$, are given by:

$$W_\alpha(\mathcal{K}_1 \cap \mathcal{K}_2) = W_\alpha(\mathcal{K}_1) + W_\alpha(\mathcal{K}_2) - W_\alpha(\mathcal{K}_1 \cup \mathcal{K}_2) \ . \tag{2}$$

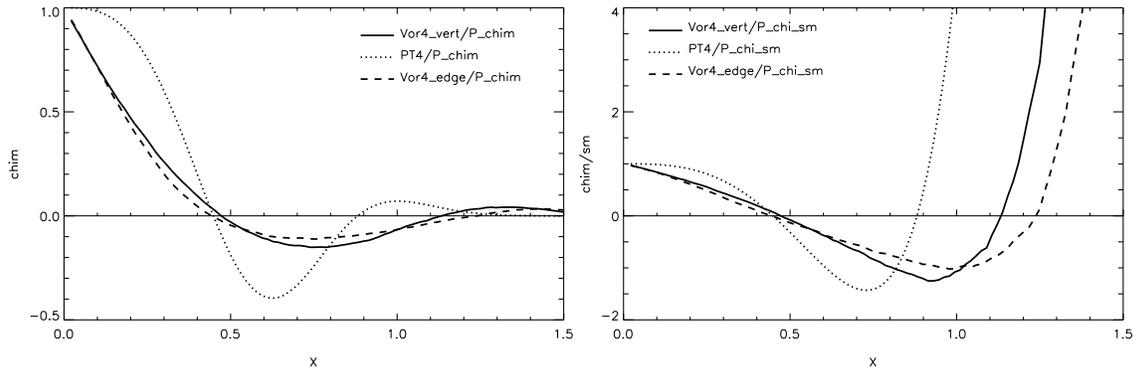

Fig. 2. Examples of the Minkowski–Functionals are depicted as a function of radius normalized by the mean interparticle separation d, x = r/d, for the two point sets of Fig.1. Left panel: the 'reduced' mean Euler–characteristic (the fourth Minkowski–Functional); right panel: the same measure but normalized by the 'reduced' total surface area of the coverage (the second Minkowski–Functional). The measures corresponding to Fig.1 (left) are shown as a full line, those corresponding to Fig.1 (right) as a dashed line, and the dotted line shows the theoretical curve of a Poissonian point set as a reference model. The correlation length of the two–point correlation function is at x = 0.55.

2. *Motion invariance* expresses the invariance of the values of these functionals with respect to translations and rotations of the body, and 3. a *continuity* property, which intuitively states that an approximation of the body by polyhedra also yields an approximate agreement of the corresponding functionals (see: Mecke *et al.* 1994).

### 2.2. Global morphology

Hadwiger's theorem asserts that the 4 Minkowski–Functionals (1) form a complete family of statistical descriptors of the convex ring, i.e., for sets formed by a finite union of intersecting convex bodies; they provide a 'fundamental system' in the space of additive, motion invariant and continuous measures.

In order to achieve a *global* characterization of point sets, we have to vary the radius $r$ and give the mean values of the Minkowski–Functionals as a function of this diagnostic scale parameter. These measures embody information from *any* order of the correlation functions (see: Mecke *et al.* 1994). Fig.2 presents two of the 'reduced' mean Minkowski measures (i.e., normalized by the measures of the unit ball and the number of sample points) for the point processes of Fig.1 with identical two–point correlation function.

### 2.3. Local morphology

The additivity property (2) is the key–element to a local characterization of patterns:

The contributions to the measures from singular intersections of the balls are calculated at each point and added up according to (2) to obtain the global average values. Consequently, we can assign to each point a set of 4 numbers together with a fixed value of the radius $r$, which we interpret as the scale of the *environment* of a point (a galaxy). This way we are able to extract subensembles from the point set with given local morphological characteristics: A galaxy in a filamentary environment will have characteristics different from that in a cluster environment.

*2.4. Finite–size effects*

A major source of uncertainty in the statistical analysis of finite samples is given by boundary effects. These finite–size effects can be handled explicitly by calculating the Minkowski–Functionals on the boundary: We again use the additivity property (2) to add up the boundary contribution, which is obtained by cutting the coverage with the body of the survey geometry (the so–called *window domain*). We currently implement geometries of all–sky survey data into the numerical program to realize the Minkowski–Functionals. This will be important for the measurement of observational samples, where boundary effects have to be seriously considered.

## 3. Comparison with 'genus statistics'

*Significance* is another consequence of additivity (2): The power of a statistical measure can be tested by going to sparse samples and by looking at the deviations from theoretically predictable results. The Minkowski–Functionals of a Poissonian sample can be calculated analytically (Mecke & Wagner 1991, Mecke *et al.* 1994).

We have performed a test which is designed similar to a test by Weinberg *et al.* (1987) of the 'genus statistics' (see: Melott 1990 for a review). The genus $g$ is related to the Euler–characteristic by $g = 1 - \chi$. However, the popular approach to calculate this measure is different: Firstly, an iso–density contour is constructed by smoothing the point set into a density field and looking at level surfaces of the density. This method involves two diagnostic parameters, the smoothing scale $\lambda$ and the density threshold $\nu$, in contrast to the single scale parameter $r$ used here. Secondly, the genus is calculated for this iso–density surface as a function of the threshold $\nu$.

Fig.3 shows an extreme test (using only 153 points to realize a Poisson process): The 'genus statistics' for a similar test (see: Weinberg *et al.* 1987, Fig.11) strongly depends on whether the concept of an iso–density surface is sensible; in other words, it needs a comparatively large smoothing length of the order of the mean separation of points in the set to recover the theoretically predicted genus curve. The mean value over six realizations then agrees within large error bounds. On the contrary, the additive calculation of the Euler–characteristic yields highly significant results which are obtained without any indirect preparation of the point set and without any statistical assumption on its distribution. (Note that the radius used as diagnostic parameter bears no simple relationship to the threshold $\nu$ in the 'genus statistics'; however both

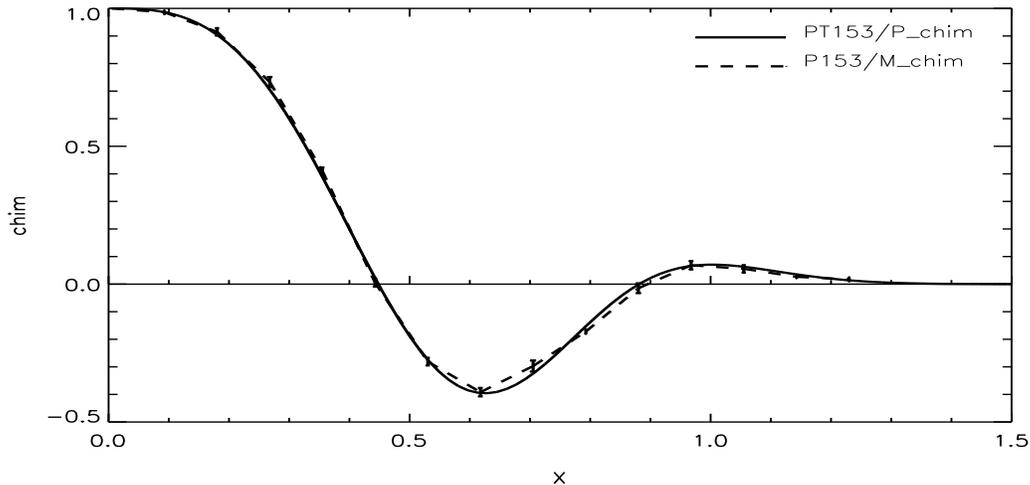

Fig. 3. A significance test is performed for the average over $6$ realizations of a Poissonian process with $153$ points (compare Weinberg *et al.* (1987) for a similar test of the 'genus statistics' for different smoothing lengths $\lambda$). Fig.3 shows the average for the 'reduced' mean Euler–characteristic (dashed line) with the same number of particles and realizations, and the same number of calculated radii of the coverage corresponding to the number of thresholds $\nu$ in the 'genus statistics'. Error bars cover $1\sigma$ dispersion from the average; to compare with the error bars in (Weinberg *et al.* 1987, Fig.11) they have to be multiplied by $\sqrt{6}$. The theoretical Poissonian is shown as a full line.

parameters measure the change of topology in either of the two methods.)

The comparison of Fig.2 (left) with Fig.2 (right) shows further that the Euler–characteristic itself is, although significant, not favorable as a discriminator of point sets: Combinations of the 4 Minkowski measures should be used to extract the full morphological information.

Still, the 'genus statistics' is a complementary tool; it is relatively insensitive to the details of non–linear evolution and biasing, and so makes the connection between present and initial topology reasonably direct. In particular, it can be employed for the characterization of density fields on a grid. Brandenberger *et al.* (1994) have recently modified this method to a 'discrete genus statistics', and Mo & Buchert (1990) have studied geometrical and topological discriminators on a grid (compare also Shandarin 1994). However, an additive and motion invariant tool is preferable: We therefore plan to investigate the Minkowski measures on a grid for the application to density or temperature fields.

## 4. From 'soft' to 'hard' clustering statistics

A complete family of morphological descriptors based on Minkowski–Functionals has been proposed for the robust characterization of large–scale structure (Mecke *et al.* 1994).

With the help of 4 significant measures we are able to globally characterize point sets with a single diagnostic scale parameter as well as to give a local morphological characterization of galaxy environments.

This novel method conceptually generalizes earlier efforts, which have put forward the Euler–characteristic as a statistical tool for the measurement of large–scale structure (compare Doroshkevich 1970 and Melott 1990).

Besides the possibility to take finite–size effects explicitly into account, we plan to exploit the *stereological* properties of Minkowski–Functionals (i.e., the extraction of three–dimensional information from low–dimensional sets such as 'pencilbeams' or two–dimensional all–sky survey data, see: Weil 1983).

A discussion remark given by Fred Bookstein (in: Feigelson & Babu 1992) can be used as a summary of the intended message of this talk: He criticized cosmologists who "emphasize *hard* physical theories of how galaxy clustering arises, but put forward mainly *soft* statistical tactics of their study". To bridge this gap he refers to the literature of image processing and stochastic geometry, and independently suggested the "hard model of Minkowski–Functionals of convex grains" as a possibly fruitful alternative (we thank the referee of our paper for pointing us to this discussion remark which I recommend to read).

The code to realize the Minkowski–Functionals can be obtained via electronic mail: tob@mpa-garching.mpg.de

## 5. Acknowledgements

I would like to thank Rien van de Weygaert for generating the point sets of Fig.1, Michael Platzöder for running the 'minkowski' program resulting in Fig.2 and Fig.3, as well as David Weinberg for discussions.